\documentclass[smallcondensed]{article}

\usepackage{blindtext} 
\usepackage{graphics}
\usepackage{graphicx}

\usepackage[sc]{mathpazo} 
\usepackage[T1]{fontenc} 
\usepackage{microtype} 

\usepackage[english]{babel} 

\usepackage[hmarginratio=1:1,top=32mm,columnsep=20pt]{geometry} 
\usepackage[hang, small,labelfont=bf,up,textfont=it,up]{caption} 
\usepackage{booktabs} 

\usepackage{lettrine} 

\usepackage{enumitem} 
\setlist[itemize]{noitemsep} 

\usepackage{abstract} 

\usepackage{titlesec} 
\renewcommand\thesection{\Roman{section}} 
\renewcommand\thesubsection{\roman{subsection}} 
\titleformat{\section}[block]{\large\scshape\centering}{\thesection.}{1em}{} 
\titleformat{\subsection}[block]{\large}{\thesubsection.}{1em}{} 

\usepackage{fancyhdr} 
\usepackage{titling} 

\usepackage{hyperref} 

\usepackage{amsmath}
\usepackage{array}
\usepackage[english]{babel}

\newcommand\Tablefontsm{\fontsize{8}{9}\selectfont}

\usepackage[flushleft]{threeparttable}
\newcolumntype{L}[1]{>{\raggedright\let\newline\\\arraybackslash\hspace{0pt}}m{#1}}
\newcolumntype{C}[1]{>{\centering\let\newline\\\arraybackslash\hspace{0pt}}m{#1}}
\newcolumntype{R}[1]{>{\raggedleft\let\newline\\\arraybackslash\hspace{0pt}}m{#1}}

\pretitle{\begin{center}\Huge\bfseries} 
\posttitle{\end{center}} 
\title{Heterogeneity in Demand and Oprimal Price Conditioning for Local Rail Transportation} 
\author{%
\textsc{Evgeniy M. Ozhegov}\thanks{Corresponding author} \\[1ex] 
\normalsize National Research University Higher School of Economics.\\Research Group for Applied Markets and Enterprises Studies. Research fellow \\ 
\normalsize \href{mailto:tos600@gmail.com}{tos600@gmail.com} 
\and 
\textsc{Alina Ozhegova} \\[1ex] 
\normalsize National Research University Higher School of Economics.\\Research Group for Applied Markets and Enterprises Studies. Junior research fellow \\ 
\normalsize \href{mailto:arbuzanakoba@gmail.com}{arbuzanakova@gmail.com}}

\date{\today} 
\renewcommand{\maketitlehookd}{%
\begin{abstract}

This paper describes the results of research project on optimal pricing for LLC "Perm Local Rail Company". In this study we propose a regression tree based approach for estimation of demand function for local rail tickets considering high degree of demand heterogeneity by various trip directions and the goals of travel. Employing detailed data on ticket sales for 5 years we estimate the parameters of demand function and reveal the significant variation in price elasticity of demand. While in average the demand is elastic by price, near a quarter of trips is characterized by weakly elastic demand. Lower elasticity of demand is correlated with lower degree of competition with other transport and inflexible frequency of travel.
\end{abstract}
}

\begin{document}

\maketitle

\section{Introduction}

Local rail transportation in Russia is an important part of public transport system, as it is often the only way for people to move between locations. At the same time, due to the large length of the tracks and the distance between large settlements, regional passenger companies cannot cover the costs of operating local rail transport by the revenue from ticket sales. For instance, the revenue of LLC Perm Local Rail Company (PPK) in 2017 amounted to 596 million rubles (near 1 mln. euro), while the company's expenses were 866 million rubles. This significant gap between revenue and expenses mostly arises due to Gornozavodsk direction, where rail transport is the only available type of transport for its residents and is covered by a subsidy from the regional budget (about 400 million rubles, about 0.4\% of the budget expenditures of the Perm Territory). \par
In order to optimize the revenue and minimize the amount of subsidies, PPK initiated an industrial research project to "develop an integrated model of tariff regulation for the local rail transportation system in the Perm Territory in the long run". This article describes the results of the study devoted to the demand for local rail transportation, which are used to obtain the optimal tariffs for transportation. \par
Local rail transport is a classic example of a "perishable" product, the specific features of which include the inability to stock a product, a limited amount of stock from the seller, low marginal and high fixed costs, and variable demand (Ly, 2012). Revenue management (RM) is a common way of maximising profits in capacity-constrained industries, such as theaters, sports events, air transport, hotel services and local rail transport. Previous papers in the field of revenue management proves that price discrimination is an effective strategy to increase the revenue (Hetrakul and Cirillo, 2013). The strategy of price discrimination is based on the idea that the price of a good is distinct for each consumer segment and must be charged in accordance with the willingness to pay, which can be revealed from the estimation of the demand function. In the absence of explicit information about the characteristics of consumers, the transportation market can be separated into submarkets for each of the possible directions. Price optimization policy can be carried out in the context of submarkets. In this case, the optimal price is determined by characteristics of a trip (price conditioning) rather than by characteristics of consumers.  \par
At the moment, the PPK pricing strategy implies price differentiation based on the trip distance only. If any other factors, rather than the trip distance, influence market capacity or price elasticity, then further tariff differentiation may lead to an increase in the company's revenue. \par
Previous papers on the demand for rail transport reveal that in addition to the price, the number of tickets sold and the elasticity of demand can also be affected by competition with other types of transport, travel goals of passengers, weather conditions, characteristics of departure and arrival stations (see, for example, Rietveld, 2000; Vuuren, 2000; Vuuren and Rietveld, 2002; Paulley \textit{et al.}, 2006; Zemp \textit{et al.}, 2011; Hetrakul and Cirillo, 2013). Thus, to estimate the demand function, we employ data on PPK ticket sales for the entire period of the company's operation available to the date of research (2012-2016, about 10 million tickets), information on the actual train schedule (about 130 thousand trains), and station characteristics within the PPK coverage area (215 stations). Characteristics of stations include information about the nearest settlements, distance from the station to the settlements, bus stop, highways and the regional center.\par
In order to account for the potential unobserved demand heterogeneity, the estimation of the demand function for local rail transportation is carried out by the statistical methods that take into account the differences in the estimates of the demand parameters among subsamples of trips. The use of fixed-effect panel data models with fixed effects on the trip date, the direction and distance of the trip allows to control  for the potential problem of unobserved demand shocks. However, estimates of price elasticity and other parameters are considered constant for all trips in the framework of fixed-effect models. One of way for modelling the demand heterogeneity is to use mixed models. However, one of the main drawbacks of such models is their non-interpretability in terms of the heterogeneity causes. Therefore, in this study along with the fixed-effects model, we use the approach of bootstrap aggregation of regression trees (Breiman, 2006), which is based on the idea of heterogeneous parameter estimates in different data subsamples.\par
Estimation of the parameters of the regression tree model allows to conclude that the demand for the average rail trip is elastic by price, which also coincides with the estimates of the conventional panel data model with fixed effects. The regression tree model also reveals that three quarters of the trips are characterized by elastic demand by price, while the demand for the rest quarter of trips is weakly elastic. At the same time, the least elastic demand mostly includes the trips between remote settlements from the regional center in the Gronozavodsk direction, where the trips are carried out mainly for work, i.e. without the ability of flexible change the frequency of travel, and in the absence of alternative transport choice. The most elastic demand is observed for the trips in the areas with high degree of competition with local and city buses, as well as the trips with the purpose of vacation and summer gardening, which have a more flexible frequency. We use these findings to suggest the optimal the pricing strategy for the PPK. The estimate of revenue increase allow to use the strategy of more deep price differentiation in the direction of revealed sources of demand heteorogeneity, such as trip distance, direction and goal of trip.\par
The structure of the paper is organized as follows. The next section describes the local transportation market in the Perm territory. The following is an overview of approaches to model the demand for transport. The fourth section describes the data used in the study. The following sections contain the research methodology and the results of the empirical analysis of demand. Then we describe the optimization problem and its main results. The last section describes the key results of the research and its limitations.   
\section{Overview of rail transport market in Perm region}

The current pricing policy of the PPK is based on the zone system. The cost of travel in any direction, regardless of the day of the week and month, depends only on the distance of the trip. The distances are divided into intervals, so-called zones. The distance for each zone varies from 10 to 15 km. The tariff monotonously grows with an increase in the distance of the trip, and the increase is not constant. This arises due to the fact that the current pricing policy already takes into account the heterogeneous demand for trips within the particular tariff zones. However, in general, the principle of decreasing payment for each additional kilometer is observed with an increase in the distance of the trip.\par
The maximum value of the full fare for a trip is set by the Regional Service for Tariffs (RST). Hence the marginal fare for a trip within the first zone in 2016 could not exceed 26 rubles (near 0.4 euro). Each year, the tariff in the first zone is set by PPK equal to the upper bound determined by the RST. The exception is 2017, when the fare for the trip within the first zone and the Perm agglomeration was set to 20 rubles in accordance with the price of trip on public transport within the city of Perm. The value of the tariff in the remaining zones is currently regulated by the RST and is set by the PPK equal to the upper bound value.
Tariff value changes annualy on January 1 of each year. The average tariff growth rate in 2012–2017 was 8.5\%, which corresponds to the average price growth rate, while the growth in individual years and zones was not uniform. \par
The current tariff policy implies the availability of single trip and round trip tickets as well as season tickets. Fare for the round trip ticket is discounted on 10\% from the total cost of two tickets (from 2017 the discount rate is 5\%). The share of the number of round trips in the total volume of tickets purchased increased from 9.5\% in 2013 to 25.1\% in 2016, with approximately the same total number of trips. Season (subscription) tickets vary by type: tickets for weekdays / weekends / all days, even / odd / all days. Also, subscriptions vary in duration. The tariff for them is set based on the expected number of trips multiplied by the full fare and the amount of the discount. The general principle of price schedule construction from the full fare single ticket tariff was the same across the observational period.\par
The current pricing policy provides tickets at reduced cost from the full fare. There is a reduced fare for children under 7 years of age (0.1\% of all tickets sold), for categories of citizens for whom federal (11\% tickets) and regional discounts (3\%) are set, as well as employees for the PPK and the Russian Railways (16\%) and soldiers (0.03\%). The total cost of discounted tickets is set on the basis of the cost of the full fare and discount set by the relevant federal or regional regulatory act. The value of the discount during the analysis period (2012-2016) remained constant. Thus, different categories of citizens are assigned a different tariff in accordance with their affiliation to one of the discount groups. At the same time, the tariff for these categories is set based on the need to provide social benefits. It does not take into account the amount of effective demand.\par
In total, there are three directions in the Perm Territory where rail transportation is carried out: Western (Perm - Vereshchagino - Balezino), Kungur (Perm - Kungur - Kordon), Gornozavodsk (Perm - Chusovoy - Evropeyskaia, Perm - Lys'va - Un', Chusovoy - Un', Perm - Ugleuralskaya - Kizel - Park-Kaliinaia). \par
The Western direction the railway is almostly not duplicated by the main highway, being at a distance of 20-40 km from it. Large settlements lying near this railway line have a weak automobile and bus service. Also a large number of stations belonging to the urban agglomeration of Perm belongs to this direction. \par
In contrast to the western, the southern direction to Kungur is almost completely duplicated by a network of high-quality roads and, as a result, is characterized by a developed bus accessibility of settlements lying on the route of a local train. Most of these stations are characterized by the location of large summer gardening settlements in their vicinity, which causes a relatively high passenger traffic in summer.\par 
Gornozavodsk direction in terms of distance from Perm is divided into two connecting branches. Both branches are characterized by relative duplication with the highways for the first 50 km in largely populated areas, areas without duplication by highways to large regional cities, as well as the connection of large regional cities with old metallurgical, mining and industrial settlements. In addition, an important feature of the direction is the location of the stations, connecting different parts of the city with faster communication with the center (Perm-1 and Perm-2) compared with the bus.
\section{Review of rail transport demand studies}

Due to certain characteristics, rail transportation is an industry producing perishable good. The properties of such industries include (Ly, 2012): a) variable and uncertain demand, which means that demand varies greatly depending on the time of day, day of the week, season, a pair of locations of the beginning and end of transportation, the purpose of the trip; b) the stock of goods cannot be inventoried for a long time, i.e. unfilled coaches and / or seats in a particular train cannot be put into stock and sold at another time. Vice versa there can be a decreased revenue from each train in the case where potential demand exceeds the capacity of train; c) relatively low marginal costs (costs for transporting one additional passenger) in relation to significant fixed costs (costs for traveling of single train in one direction); d) the product (ticket) can be sold in advance, but not earlier than 2 weeks before the date of travel. Finally, consumers are heterogeneous and can be divided into segments by age, income, purpose of the trip, trip time within the day, trip day within the week.\par 
However, the local rail industry has features that are not common to other similar industries, such as air transport and intercity rail transport. Local rail transportation is characterized by the presence of intermediate stops between the points of the beginning and end of the route of the train (as opposed to air transport). Therefore, each pair of stops should be a analyzed separately to determine the optimal price of transportation between these two stops and the potential loss of revenue from inefficient pricing. The presence of intermediate stations also allows the passenger to split the trip into separate segments and buy several tickets for short trips instead of one for a long one, if it is more worth for a passenger. This fact also does not allow to perform optimization of tariffs for each pair of stations, regardless of the tariffs for other trips due to the interrelated demand for travel.\par
The demand for local rail transportation is characterized by a conventionally unlimited capacity, unlike, for example, air transport and long-distance trains. Since tickets are not sold on the seats in a particular train, the number of passengers actually transported may differ from the standard capacity of the coaches. However, the expectation of the ratio of the number of passengers to the number of coaches may affect the willingness of passengers to use particular trains or rail transport in general. The capacity of transportation for a particular trip is limited and depends not only on the capacity of a particular train, but also on the train's fullness due to passengers traveling in the same train, but along a different route (for example, in the same direction, but with different stops as beginning and ending). This fact does not allow for the optimization of tariffs independently for individual pairs of stations due to the interrelated supply of goods also.\par
It is also hard to estimate the actual demand for a particular train (as opposed to inter-city passenger transport by rail) due to the lack of a ticket registration procedure, the lack of an explicit linkage of a ticket to a specific place and train. The situation is complicated by the availability of round trip tickets and season tickets, which can be used at any time and train. 
Previous studies prove that for goods and services belonging to the category of perishable goods, the strategy of price differentiation depending on the characteristics of demand (price discrimination) is an effective pricing strategy to increase the revenue (Ly, 2012). At the same time, price and revenue management is usually based on the assumption that different consumers are willing to pay different prices for the purchase of a product. In this case, the optimal pricing strategy is to charge her reserve price for each consumer, or to set a price based on the average willingness to pay related to a certain consumer segment. \par
In traditional models, consumer choice is based on the notion of a reserve price, the estimate of which can be obtained from empirical data or from consumer characteristics by estimating the variation of parameters across a population. The distribution of reserve prices is traditionally modeled using the aggregate demand function (Talluri and van Ryzin, 2004a). In later works, consumer choice models are calibrated on individual data and used to optimize revenue and pricing (Talluri and van Ryzin, 2004b; Vulcano \textit{et al.}, 2010; Chaneton and Vulcano, 2011; Newman \textit{et al.}, 2012). Among such models, probabilistic models based on utility theory with random coefficients (Ben-Akiva and Lerman, 1985) are used to model preferences of a heterogeneous population. Within this approach, consumers can be segmented based on their socio-demographic characteristics (Cherchi and Ortúzar, 2003), divided into groups with similar preferences or reactions to price changes (Carrier, 2003; Hetrakul and Cirillo, 2013), or have a set of preferences derived from a random distribution. In rail transport demand models based on deterministic heterogeneity (multiple or cluster choice models, multinomial and nested logit models), consumers are usually segmented by income and / or travel goals. In the case unobservable segments, it is necessary to use more advanced approaches. Thus, in the model of choice with latent consumer segments (latent class discrete choice models), consumers of each segment are considered to have the same preferences parameters. In addition to the preferences parameters for each segment, the probability of the consumer belonging to each of the segments is also estimated (Talluri, van Ryzin, 2004a). \par
In recent years, researchers from various disciplines have increasingly used mixed logit models to analyze complex transport industry phenomena, such as analyzing the value of travel time (Greene \textit{et al.}, 2006), airport selection (Hess and Polak, 2005) and the air company (Carrier, 2003), the choice of mode of transport for the trip (Brownstone \textit{et al.}, 2000; Hess \textit{et al.}, 2006), as well as the effect of transport fullness on demand (Bhat and Castelar, 2002). Mixed models have a huge advantage over other models, because allow to obtain an estimate of the variation in passenger sensitivity to various impacts, which significantly increases the predictive accuracy of the transportation demand model (Hetrakul and Cirillo, 2013). In addition, machine learning models that divide the sample into separate homogeneous subsamples are also used to estimate the demand for a heterogeneous population (Bajari \textit{et al.}, 2015). This approach allows to estimate the price elasticity of demand and the size of the market for each subsample of observations. In this case, subsamples can be separated on the basis of travel directions, zones, stations of departure and arrival, etc.\par
In previous studies on the estimation of the demand function for rail transportation, the price of the ticket and the characteristics of the train (speed, fullness, convenience, number of cars) are revealed as main factors of demand. Some trains offer coaches with various level comfort, which should be incorporated into the demand model (Whelan and Johnson, 2004).The demand for local transportation largely depends on the availability of this type of transport for passengers; for this purpose, the model includes the distance from the house to the train station (Wardman, Lythgoe and Whelan, 2007), railway station infrastructure indicators, for example, integration with urban and bicycle transport . The market capacity of a particular route depends largely on the development indicators of the settlement near the station (Zemp \textit{et al.}, 2011). At the same time, it is necessary to take into account the presence of substitutes for rail transportation, for which it is necessary to include in the model the availability of alternative types of transport and their characteristics (travel time, comfort) (Morikawa, McFadden and Ben-Akiva, 2002).As previously discussed, the demand for passenger transport changes significantly over time, then it is necessary to take into account the season (time of day, day of the week, month of the year, holidays). Since local railway transport can be used by travellers with various goals, it is necessary to take into account the purpose of the trip, for example, a home-work trip or a home-garden trip. While travel time influences the demand for trips of the first type, weather conditions affects the second type trips (Wheelan \textit{et al.}, 2008). In addition, the price elasticity of demand is affected by the actual value of the tariff, which can vary depending on whether the passenger belongs to a category of discount (retired, student, employees of the PPK and the Russian Railways) (Wheelan \textit{et al.}, 2008).

\section{Data description}

\subsection{Primary data}

The data for the study include information on tickets sold for 2012-2016, the actual train schedule, current tariffs and were taken from the sales system of the PPK. Data on the actual schedule of trains contain information on the train number, station of departure and arrival, intermediate stations on the route, as well as day and time of the day. For each train there is an information about the number of coaches and the type of train (fast / not fast). In 5 years, about 130 thousand trains entered the route. \par
There are 215 stations in the PPK coverage area, which leads to the fact that there are 215 $\times$ 215 possible routes for passengers to follow. However, some of these routes are not actually used by passengers, since the stations are located in different directions. In addition, there is no passenger movement between certain stations due to the lack of demand, which will also lead to zero observed demand for the route. Thus, out of 215 $\times$ 215 possible pairs in the data, only movement in 6593 routes is observed. For each of these route options, we collected the information about the stations of departure and arrival, as well as the distance between them. \par
The observed demand for the route is a number of tickets purchased between a pair of stations on a specific date. In five years, a total of about 10 million tickets were sold. For each route, we know the direction of trip, the station of departure and arrival. For each ticket we know the type of ticket (single, round or season trip), as well as the category of discount.  \par
In addition, information about tariffs is known by the type of ticket, date of travel and distance between stations. In addition to data taken from the sales system, we collected the information information about each station. For each station we observe the size of the two nearest settlements, the distance to the two nearest settlements, the presence of a bus stop within the vicinity of the closest settlement, the distance to the bus stop, the distance to the highway, the presence of garden plots near the station and geographical coordinates of station. \par
The obtained data on ticket sales was aggregated to the level of each pair of stations, the month of the year and the type of ticket. The key indicators in such a data structure are the number of tickets sold of a particular type for a particular trip in a particular month, the real (deflated) fare for a given type of ticket, the direction and distance of the trip (in kilometers and zones), as well as the characteristics of the stations where the trip begins and ends. Further, these data were used to estimate the demand function.

\subsection{Preliminary data analysis}

A preliminary analysis of ticket sales data allows to formulate hypotheses regarding the function of demand for rail transportation. As mentioned above, the transoprtation is carried out in three main directions. Gornozavodsk direction is the main in terms of the number of tickets sold (about 50\%) (see Table 1). The trips in the subdirections to Chusovoy and Kizel are about the same proportion. It is worth noting that about a third of the direction tickets are sold for trips within the urban agglomeration of Perm. In general, the Gornozavodsck direction in terms of the growth rate of the number of tickets corresponds to the general sample of tickets, with faster growth in the subdirection to Chusovoy and Lysva and a lower rate in the direction Chusovskaya - Evropeyskaya and Chusovskaya - Un'. 
The western direction accounts for about a third of the total number of tickets sold, of which about a quarter are trips within the agglomeration area. The growth rate of the number of tickets is similar to the overall growth rate of the number of tickets sold with a significantly higher growth rate of tickets for trips within the agglomeration. \par
	
	\begin{table}%
	\caption{Shares and growth rates for sold tickets by trip directions}
	\label{tab:01}
	\Tablefontsm
	\begin{minipage}{\columnwidth}
		\begin{center}
	\begin{tabular}{ L{1.0 cm}  C{1.1 cm} C{1.5 cm}  C{1.1 cm} C{1.5 cm} C{1.1 cm} C{1.5 cm} C{1.1 cm} C{1.5 cm} }
					\noalign{\smallskip}\hline\hline\noalign{\smallskip}

				Year &	\multicolumn{4}{c}{Western}		&  \multicolumn{2}{c}{Kungur} 	&  \multicolumn{2}{c}{Agglomeration}	  \\
				 \hline\noalign{\smallskip}
				 &	\multicolumn{2}{c}{ Total}	& \multicolumn{2}{c}{Incl.  Agglomeration} &  \multicolumn{2}{c}{ Total} & \multicolumn{2}{c}{Total}	  \\
				 \hline\noalign{\smallskip}
				& Number & Growth, \% & Number & Growth, \% & Number & Growth, \% & Number & Growth, \% \\
				\noalign{\smallskip}\hline\noalign{\smallskip}

2013 &	514382	 & &	106697	 & &	242235	  &&		385072  &\\			
2014 &	578336	 &12.4 &	146641 &	37.4 &	285495 &	17.9	 &	493979 &	28.3	\\	
2015 &	572233 &	-1.1 &	165266 &	12.7 &	307298 &	7.6	 &	557988	 & 13.0\\	
2016 &	543398  &	-5.0  &	150091  &	-9.2  &	301267  &	-2.0	  &	557307  &	-0.1 \\		
\hline\noalign{\smallskip}
Share  &	32.4\%  & &				16.7\%	 & &	29.3\% & & 46.7\%		\\
	\noalign{\smallskip}\hline\noalign{\smallskip}
                     & \multicolumn{8}{c}{Gornozavodsk}	\\
                     \hline\noalign{\smallskip}
						 &	\multicolumn{2}{c}{Perm - Chusovoy,}		& \multicolumn{2}{c}{Perm - Kizel} & \multicolumn{2}{c}{Chusovoy - Evrop.,} & \multicolumn{2}{c}{Incl. Agglomeration}	  \\
			 & \multicolumn{2}{c}{Perm - Lys'va } && & \multicolumn{2}{c}{Chusovoy - Un'}	  \\
			 \hline\noalign{\smallskip}
				& Number & Growth, \% & Number & Growth, \% & Number & Growth, \% & Number & Growth, \% \\
				\noalign{\smallskip}\hline\noalign{\smallskip}

2013 &	800198	 & &	438052 & &		461404 & &		116915	 	\\
2014 &	892176 &	11.5 &	490372	 & 11.9	 & 561734 &	21.7 &	109075 &	-6.7	\\
2015  &	882121 &	-1.1 &	539143 &	9.9 &	591725 &	5.3 &	93887 &	-13.9	\\
2016 &	898643 &	1.9 &	558752 &	3.6 &	590247 &	-0.2 &	98713 &	5.1 \\
\hline\noalign{\smallskip}
Share &	50.9\%	\\
\noalign{\smallskip}\hline\noalign{\smallskip}
			\end{tabular}
		\end{center}
	\end{minipage}
\end{table}

Kungur direction is characterized by the lowest share of tickets sold among all directions. The key factors here are the lack of a large number of stations within the city lying on the direction and high competition with local bus transportation. At the same time, Kungur direction shows the highest average growth rate of the number of tickets sold.  \par
Due to the uneven location of large settlements along the route of local trains, as well as different growth rates of the number of tickets sold in various directions, the growth rates of the number of tickets sold by various tariff zones differ significantly. Table 2 shows the number of tickets sold and the growth rate of the number of tickets by years divided by tariff zones. Thus, the largest share of tickets is sold for the shortest trips (first two zones, up to 25 km). It is dominated by trips inside the agglomeration and trips to the vicinity of large regional centers (Vereshchagino and Kungur). Also, a high proportion of trips belongs to the 12 zone (the dominant trips Perm-2 - Vereshchagino and Perm-2 - Kishert), 10 zone (the dominant trip Perm-2 - Kungur) and 15 zone (Perm-2 - Ugleuralskaya). Further, approximately the same and low share have trips over medium distances  (three to nine zones, from 26 to 95 km), which mainly include trips from Perm to relatively small settlements located from Perm, respectively, to Vereshchagino, Kungur, Chusovoy and Ugleuralskaya. These trips also included trips between large regional centers (for example, Vereshchagino-Kez, Chusovoy-Biser, Kizel-Yaiva).  \par
	
	\begin{table}%
	\caption{Number of tickets sold and its growth rate by zones}
	\label{tab:02}
	\Tablefontsm
	\begin{minipage}{\columnwidth}
		\begin{center}
	\begin{tabular}{ L{1.1 cm} C{1.6 cm} C{1.6 cm} C{1.6 cm} C{1.6 cm} C{2.0 cm} C{2.0 cm} }
\noalign{\smallskip}\hline\hline\noalign{\smallskip}

Zone & 2013 & 2014 & 2015 & 2016 & Zone share & Growth rate	\\

				\noalign{\smallskip}\hline\noalign{\smallskip}
1 &	226719 &		230355 &		216270 &		215338 &			12.9\%  &		-1.7\%\\
2 &		347242 &		480816 &		551699 &		553223 &			29.5\% &		16.8\%\\
3 &		109417 &		123877 &		98214 &		98252 &			5.9\% &		-3.5\%\\
4 &		57624 &		56637 &		48934 &		49035 &			3.0\% &		-5.2\%\\
5 &		73961 &		78079 &		78707 &		78522 &			4.6\% &		2.0\%\\
6 &		79985 &		85542 &		82298 &		79765 &			4.8\% &		-0.1\%\\
7 &		21353 &		24715 &		24193 &		23523 &			1.4\% &		3.3\% \\
8 &		50435 &		53822 &		51884 &		50495 &			3.0\% &		0.0\% \\
9 &		58807 &		58117 &		56517 &		56035 &			3.4\% &		-1.6\%\\
10 &		83548 &		108174 &		123607 &		128988 &			6.8\% &		15.6\% \\
11 &		31705 &		32211 &		28838 &		28056 &			1.7\% &		-4.0\%\\
12 &		163480 &		180486 &		176461 &		166874	 &		10.1\%	 &	0.7\%\\
13 &		65422 &		55172 &		48377 &		48492 &			3.1\%	 &	-9.5\%\\
14 &		59128 &		66580 &		63707 &		60562 &			3.6\% &		0.8\%\\
15 &		88652 &		113998 &		101704 &		96719 &			5.7\%  &		2.9\% \\
16  &		16996  &		7330  &		8686  &		  7768  &			0.6\%  &		-23.0\% \\
\noalign{\smallskip}\hline\noalign{\smallskip}

		\end{tabular}
		\end{center}
	\end{minipage}
\end{table}

The most rare are long distance trips. As a rule, these are trips from Perm to the most remote parts of the region, located along the route of large regional centers. \par
There is a constant increase in the number of tickets sold to the second zone, while a typical trip here is a trip within the Perm agglomeration area or a trip to Perm / from Perm from /to the nearest suburbs, for example, Overyata. Also, a positive dynamics is demonstrated by trips to the 10th zone (mainly the Perm-2 - Kungur trip). The remaining zones are characterized by a relatively stable number of tickets sold. \par
It is worth noting that certain tariff zone has very uneven distribution of  of the number of tickets sold across trip directions. This fact is due, first of all, to the uneven location of large settlements relative to Perm, as well as the different density of settlements at certain directions. Table 3 represents the distribution of the number of tickets sold separately by travel directions and tariff zones. \par
	
	\begin{table}%
	\caption{Distribution of sold tickets across zones and directions}
	\label{tab:03}
	\Tablefontsm
	\begin{minipage}{\columnwidth}
		\begin{center}
	\begin{tabular}{ L{1.0 cm} C{1.7 cm} C{1.7 cm} C{1.7 cm} C{1.8 cm} C{1.8 cm} C{1.8 cm} }
\noalign{\smallskip}\hline\hline\noalign{\smallskip}

Zone & \multicolumn{3}{c}{Number of tickets} & \multicolumn{3}{c}{Share within zone}	\\
& Western & Kungur & Gornozavodsk & Western & Kungur & Gornozavodsk \\

				\noalign{\smallskip}\hline\noalign{\smallskip}
1 &	50269 &		10159 &		161742 &		22.6\% &		4.6\% &		72.8\%\\
2 &		125183 &		39084 &		319674 &		25.9\% &		8.1\% &		66.1\%\\
3 &		44577 &		15369 &		47493 &		41.5\% &		14.3\% &		44.2\%\\
4 &		26257 &		9297 &		17502 &		49.5\% &		17.5\% &		33.0\%\\
5 &		38186 &		20543 &		18648 &		49.4\% &		26.5\% &		24.1\% \\
6 &		39403 &		8668 &		33825 &		48.1\% &		10.6\% &		41.3\%\\
7 &		8695 &		8171 &		6868 &		36.6\% &		34.4\% &		28.9\% \\	
8 &		10087 &		8735 &		32836 &		19.5\% &		16.9\% &		63.6\%\\
9 &		40075 &		7665 &		9711 &		69.8\% &		13.3\% &		16.9\%\\
10 &		6172 &		94300 &		10606 &		5.6\% &		84.9\% &		9.5\%\\
11 &		14316 &		3450 &		12435 &		47.4\% &		11.4\% &		41.2\%\\
12 &		121064 &		34913 &		15847 &		70.5\% &		20.3\% &		9.2\%\\
13 &		659 &		3491 &		50215 &		1.2\% &		6.4\% &		92.4\%\\
14 &		5840 &		9350 &		47303 &		9.3\% &		15.0\% &		75.7\%\\
15 &		21299 &		2584 &		76384 &		21.2\% &		2.6\% &		76.2\%\\
16 &	&		8287 &		1908 &		0.0\% &		81.3\% &		18.7\% \\	
17 &		&&		5280 & 		0.0\% &		0.0\% &		100.0\%\\
\noalign{\smallskip}\hline\noalign{\smallskip}

		\end{tabular}
		\end{center}
	\end{minipage}
\end{table}

Thus, the Gornozavodsk direction is characterized by the highest share of tickets sold in zones 1 and 2 due to the large number of trips within the city agglomeration. The western direction is the leader in trips within 3-7, 9 and 12 zones (trips from Perm to the stations in front of Grigorievskaya, as well as Perm - Mendeleevo, Perm - Vereshchagino). At the same time, within the 8th zone, the main contribution is made by the Gornozavodsk direction (the Perm - Komarikhinskaya trip). A similar situation is observed for the 13th zone (Perm - Chusovoy), the 14th (Perm - Lys'va), the 15th (Perm - Ugleuralskaya). While Kungur direction dominates in the 10th and 12th zones (Perm - Kungur and Perm - Kishert trips). \par
Obviously, the zone tariff regulation is not optimal due to the fact that each zone may contain quite a lot of tickets with trips that are not the same in terms of the demand capacity and sensitivity, which may vary according to the purpose of the trip. Thus, the analysis of the dominating trips in each direction and in each tariff zone allows to conclude that quite similar trips with large shares of tickets sold can be put into one zone (for example, trips from Perm-2 station to Chaykovskaya on western direction, Kukushtan station on the Kungur direction and Div'ya station on the Gornozavodsk direction). All of them are located at approximately the same distance from Perm, are large garden villages and have approximately comparable car accessibility with a slightly better access to Kukushtan by bus.\par
Also, trips with significantly different goals and, as a result, different capacity of the trip market and price elasticity of demand can be put into the same zone (for example, the Shnyry - Kez trip as third zone trip  in the western direction, the moderate distance trip between large settlements; Perm-2 - Mulyanka in the Kungur direction, a short trip to holiday village of medium size; and Perm-2 - Golovanovo, a trip within the agglomaration in the Gornozavodsk direction).\par
Other zones, on the contrary, may have one dominant trip in one of the directions (zones 8, 10, 12-15) with a very low proportion of other trips. In this case, the optimal price for a trip to zone 8 (Perm - Komarikhinskaya) may be higher than the optimal price for a trip to zone 10 (Perm - Kungur) due to the lower elasticity of demand and competition from alternative means of transport. Thus, the analysis of the dominant trips within individual zones and directions allows to formulate the hypothesis that the zone pricing that is not considering the type of trip and the characteristics of specific departure and arrival stations, may be not optimal.\par 
For the purpose of optimal pricing, one should estimate the capacity of each local market of a separate route and the price elasticity of demand for each route, describe the trips depending on the purpose, the direction of the trip and the characteristics of the starting and ending points of the route, and determine the optimal tariffs for routes depending on the purpose of the trip and direction. Typical travel goals include regular daily trips to work (trips inside Perm, short trips between large regional centers and small settlements), weekly trips on weekends (trips to Perm and from Perm to large and medium-sized settlements), single trips on day off / weekend (trips to the countryside, trips to Perm and major settlements) and other trips. Thus, the main parameters characterizing the purpose of the trip may be the day of the week on which the trip is made, the month of the year, the size of the settlement at the beginning and end of the trip, as well as the type of ticket.\par
All settlements near the stations were divided into 5 groups by the size (Table 4): no settlement (within 5 km to the station); there is a small settlement (total number of residents living within 5 km to the station is no more than 100 people); there is a medium-sized settlement (the total number of residents within 5 km to the station is from 100 to 1000 people); there is a large settlement (the total number of residents within 5 km to the station is from 1000 to 10,000 people); there is a huge settlement (total the number of residents within 5 km to the station is above 10,000 people). \par
	
	\begin{table}%
	\caption{Distribution of sold tickets across size of settlements}
	\label{tab:04}
	\Tablefontsm
	\begin{minipage}{\columnwidth}
		\begin{center}
	\begin{tabular}{ L{2.0 cm} C{1.5 cm} C{1.5 cm} C{1.5 cm} C{1.5 cm} C{1.5 cm} C{1.5 cm} }
\noalign{\smallskip}\hline\hline\noalign{\smallskip}

From/ Where & No settl. & Small & Middle & Large & Huge & Total \\

				\noalign{\smallskip}\hline\noalign{\smallskip}

No settlment & 	0.00\% & 	0.00\% & 	0.01\% & 	0.08\% & 	0.49\% & 	0.58\%\\
Small & 	0.00\% & 	0.00\% & 	0.11\% & 	1.28\% & 	4.86\% & 	6.25\%\\
Middle & 	0.00\% & 	0.01\% & 	0.20\% & 	1.84\% & 	12.17\% & 	14.22\% \\
Large  & 	0.00\%  & 	0.01\% & 	0.47\% & 	6.27\% & 	26.50\% & 	33.24\%\\
Huge & 	0.00\% & 	0.02\% & 	1.30\% & 	11.62\% & 	32.77\% & 	45.71\%\\
Total & 	0.00\% & 	0.04\% & 	2.09\% & 	21.08\% & 	76.79\%	 \\
\noalign{\smallskip}\hline\noalign{\smallskip}

		\end{tabular}
		\end{center}
	\end{minipage}
\end{table}

The number of tickets sold is positively correlated both with the size of the settlement located near the station where the route began, and with the size of the settlement located at the end of the route. About a third of the tickets (32\%) are sold on trips between two stations that lie inside major settlements (including the Perm agglomeration, which accounts for about 16\% of all tickets sold). Also, about a third of tickets (38\%) are sold for travel between huge and large settlements. Significantly fewer tickets are sold on routes, one of the ends of which is a large settlement, the other is an average one (13\%) or a small one (5\%), and also where a large settlement is located at both ends (6\%). The share of travel which falls on the remaining types of station pairs does not exceed 5\%.\par 
The average size of the settlement for the trip from which the ticket is bought is less than the point of arrival. This fact characterizes the relatively large share of single tickets on trips between large stations as compared to trips, where at least one of the ends of the route is not a large station.\par
Thus, an analysis of the number of tickets sold in 2012-2016 allows to conclude that passenger traffic has a high degree of heterogeneity in different train directions. All trips can be divided into several key types: a short trip along the agglomeration (a trip on a weekday, single ticket); a short and medium distance trip around a large settlement (excluding Perm) (a trip both on weekends and on weekdays, equal the share of single and return tickets by days); a trip from Perm to a large settlement or \textit{vice versa} (long range, a large proportion of round trips on weekends); a trip from Perm to a small (summer) village (medium range, a large proportion of round trips over the weekend).

\section{Methodology}

We estimate a demand model using bootstrap aggregation of regression trees (bagging prediction) (Breiman, 1996), which takes into account the heterogeneity in preferences across various types of trips and the demand response to price. The bagging model is an aggregation of a family of regression tree models.  Estimation of each model is based on random subsample of data. Aggregation of predictions obtained from individual trees avoids the problem of the presence of influential observations. The regression tree model allows to obtain estimates of the elasticity of demand by price, as well as the influence of other factors on demand, that differ for different subgroups of observations (station pairs, trip directions, trip types) using algorithmic partitioning of trips' sample into subsamples.\par
The aggregate demand model predicts an aggregated choice of passengers whether or not to use local rail transport on a particular route, defined as a pair of stations. Thus, for each month and for each pair of stops we predict a number of tickets sold taking into account the price of the ticket (real fare), the characteristics of the locations of the beginning and end of the route, the month and year of a trip and the attributes of the trip. The aggregate demand model can be written as:
\begin{equation}
    q_{jkt} = f(x_{kjt},p_{kjt},z_t | \beta,\alpha, \gamma) + \epsilon_{kjt} = x_{kjt}\beta + \alpha p_{jkt} + z_t \gamma_{jk} + \epsilon_{jkt}
\end{equation}

\noindent where\\
$q_{jkt}$ is a log of the number of tickets sold from station $j$ to station $k$ in the month of travel $t$;\\
$x_{kjt}$ are attributes of the trip between stations $j$ and $k$ (distance between stations and the number of zones covered, direction of the trip, characteristics of the stations at the beginning and end of the route);\\
$\beta$ is a sensitivity of aggregate demand to trip attributes;\\
$ p_{jkt}$ is a log of real full fare for a trip from station $j$ to station $k$ in a month $t$;\\
$\alpha$ is an elasticity of aggregate demand to the price of the trip;\\
$z_t$ are time characteristics of the trip in a month $t$, i.e. month within the year and year of travel;\\
$\gamma_{jk}$ are parameters of seasonal sensitivity of demand to the trip between stations $j$ and $k$;\\
$\epsilon_{jkt}$ is an idiosyncratic shock of travel demand from station $j$ to station $k$ in a month $t$.\\

	A regression tree is a collection of rules that determine the parameters values of a regression function, $\theta = \{\beta, \alpha, \gamma\}$. Tree-based methods partition the characteristic space into a series of hyper-cubes, and fits effects to each partition depend on the value of right-hand side variables, $w = \{x,p,z\}$. Trees are characterized by a hierarchical series of nodes,
	with a decision rule associated at each node. Following (Bajari \textit{et al.}, 2015), define a pair of half-planes:
	
	\begin{equation}
R_1(m; s) = \{w | w_m\leq s\} 
		\end{equation}
	\begin{equation} 
	R_2(m; s) = \{w|w_m > s\}	\nonumber
 	\end{equation}		

\noindent where $m$ indexing a splitting variable and $s$ is a split point (threshold). Starting with the base node at the top of the tree, the rule for that node is formed by the following optimization problem:

	\begin{equation}
\min\limits_{m,s}[\min\limits_{\theta_1}\sum\limits_{j,k,t: w_{jkt} \in R_1(m,s)}L(q_{jkt} - f(w_{jkt}|\theta_1)) + \min\limits_{\theta_2}\sum\limits_{j,k,t: w_{jkt} \in R_2(m,s)}L(q_{jkt} - f(w_{jkt}|\theta_2))]
	\end{equation}

The inner optimization is solved by setting $\theta$ optimal according to prespecified loss function $L$ and regression function $f$. For ordinary regression tree $L$ is a squared function and $f$ is a linear function of $W$ with parameters $\theta$ as defined in Eq. 1. The outer optimization problem is a problem of finding an optimal splitting point $s$ for each possible splitting variable and then choosing a variable $W$ to split by. Once the splitting variable and point are found, the same procedure is then performed on each resulting partitioning, finally giving a partition of trip characteristics space.\par
In the limit, each value of $w\in W$ is assigned to value of $q=f(w)$, which is a perfect reconstruction of the underlying function $f$ for in-sample prediction. In practice, we are interested in out-of-sample prediction. Therefore, the tree is expanded until a value of loss function for out-of-sample data falls. Often, tree is grown until a specific number of splits or a minimal number of observations in subsamples is achieved.\par
The literature has proposed several variations on the regression tree estimators to obtain $"$honest$"$ prediction and predicted values robust to influential observations. One is bagging (Breiman 1996), which uses resampling and model averaging to obtain a predictor. The idea is to sample the data with replacement $B$ times, train a regression tree on each resampled set of data, and then predict the outcome at each $w \in W$ through a simple average of the predictions under each of the $B$ trees. We use the same idea of resampling, taking each time random subset of observations to train the model and predict the values of $q$ for remain observations. Calibration of 2000 regression trees with 75\% of random observations for model training at each resample expectedly gives 500 out-of-sample predictions for each observation which we average to obtain $"$honest$"$ prediction.

\section{Results}

The goal of the study is to estimate the sensitivity of demand to changes in tariff, as well as reveal the sources of heterogeneity of the price elasticity parameter. We test various specifications of the demand equation to check the robustness of the estimates. Firstly, we estimate Eq. (1) by the least squares method, i.e. not taking into account potential differences in the sensitivity of demand to tariff changes. This approach allows to check the robustness of the average price elasticity parameter to the choice of control variables that may also affect the demand for local rail transport. \par
Estimates for the various specifications of Eq. (1) using the least squares method are presented in Table 5. Each of the specifications of Eq. (1) is an equation explaining the variation of the logarithm of the number of single tickets sold for a trip in a particular month for a particular pair of stations for the full fare. The table shows that when controlling for variables which are correlated with price (fixed effects on tariff zones, year and month of travel), the estimate of price elasticity for an average individual for different specifications varies from - 1.8 to -2.0, which indicates elastic demand. This fact suggests that the revenue from trips of the average consumer buying single full fare ticket will increase with a decrease in the real tariff.\par
	\begin{table}%
	\caption{Estimates of price elasticity for various specifications}
	\label{tab:05}
	\Tablefontsm
	\begin{minipage}{\columnwidth}
		\begin{center}
	\begin{tabular}{ L{2.6 cm} C{2.3 cm} C{2.3 cm} C{2.3 cm} C{2.3 cm} }
\noalign{\smallskip}\hline\hline\noalign{\smallskip}

 & (I) & (II) & (III) & (IV)\\  
				\noalign{\smallskip}\hline\noalign{\smallskip}
Log. of real tariff & -0.180$^{***}$ &	-1.797$^{***}$ &	-1.603$^{*}$ &	-1.981$^{*}$\\
&	(0.012) &	(0.505)	& (0.884) &	(0.791)\\
\noalign{\smallskip}\hline\noalign{\smallskip}
\multicolumn{5}{l}{Control variables:}\\
Year, month &	- &	+ &	+ &	+\\
Tariff zones	 & - &	+ &	+ &	+\\
Directions &	- &	- &	+ &	+ \\
Characteristics of trip and stations  &	- &	-	 &- &	+\\
\noalign{\smallskip}\hline\noalign{\smallskip}
Number of obs. &	48311 &	48311 &	48311 &	46266\\
Number of params. &	2 &	34 &	40 &	67\\
$R^2$  &	0.004 &	0.045 &	0.090 &	0.300\\
$R_{adj}^2$  &	0.004 &	0.045 &	0.089 &	0.299\\
\noalign{\smallskip}\hline\noalign{\smallskip}
\multicolumn{5}{l}{Notes: Table cells represent parameter estimates by OLS,}\\
\multicolumn{5}{l}{Standard errors in parenthesis.}\\
\multicolumn{5}{l}{Signifinance levels are $^{*}$ $p < 0.1$, $^{**}$ $p < 0.05$, $^{***}$ $p < 0.01$.}

		\end{tabular}
		\end{center}
	\end{minipage}
\end{table}

Comparing the parameters of price elasticity in the specifications (I) and (II) suggests that when ommiting variables correlated with the price, the price elasticity estimate is turned to be biased. At the same time, the introduction of other control variables (trip directions and characteristics of stations at the beginning and end of the trip) does not statistically affect the price parameter estimate. However, the estimate of the explained variation of demand increases both when the time / seasonal component (month of the year and year of the trip) is included in the model, and when the trip direction and station characteristics are included. This suggests that demand is influenced by seasonal factors. In addition, the composition of the inhabitants of settlements located near the end points of the route, as well as differences in travel directions are also correlated with the demand for a particular trip route.\par
To check the robustness of the results we estimate the parameters of the demand model for various types of tariffs: the full fare for a single and return trip, a children's fare, a fare for the PPK and the Russian Railways employees, a fare for federal and regional discount recipients. Models for subsamples of students and military were not estimated due to the insufficient number of observations. The results of model estimation for various subsamples by the tariff type are presented in Table 6. \par 
	\begin{table}%
	\caption{Estimates of price elasticity for various types of tariffs}
	\label{tab:06}
	\Tablefontsm
	\begin{minipage}{\columnwidth}
		\begin{center}
	\begin{tabular}{ L{2.6 cm} C{1.5 cm} C{1.5 cm} C{1.5 cm} C{1.5 cm} C{1.5 cm} C{1.5 cm} }
\noalign{\smallskip}\hline\hline\noalign{\smallskip}

 & (V) & (VI) & (VII) & (VIII) & (IX) & (X) \\
 & \multicolumn{2}{c}{Full fare} & Children & PPK empl. & Fed. disc. & Reg. disc.\\
 & Single & Return \\
				\noalign{\smallskip}\hline\noalign{\smallskip}
Log. of real tariff & -1.944$^{**}$ &	-1.936$^{*}$ &	-0.822 &	-0.543 &	-1.961$^{***}$ &	-2.029$^{*}$\\
&	(0.797) &	(1.126) &	(0.741) &	(0.540) &	(0.622) &	(1.161)\\

\noalign{\smallskip}\hline\noalign{\smallskip}
Number of obs. & 44114 & 23530 & 19560 & 64924 & 48373 & 16945\\
Number of params. &	67 &	67 &	67 &	67 &	67 &	67\\
$R^2$  &	0.313 &	0.261 &	0.188 &	0.189 &	0.191 &	0.252\\
$R_{adj}^2$  &	0.312 &	0.259 &	0.185 &	0.188 &	0.189 &	0.249\\
\noalign{\smallskip}\hline\noalign{\smallskip}
\multicolumn{5}{l}{Notes: All specifications contains the set of variables as in (IV),}\\
\multicolumn{5}{l}{Table cells represent parameter estimates by OLS,}\\
\multicolumn{5}{l}{Standard errors in parenthesis.}\\
\multicolumn{5}{l}{Signifinance levels are $^{*}$ $p < 0.1$, $^{**}$ $p < 0.05$, $^{***}$ $p < 0.01$.}

		\end{tabular}
		\end{center}
	\end{minipage}
\end{table}

The results show that the average passenger demand in the categories of passengers traveling under the full fare for single and return ticket, federal and regional discount recipients are not statistically different. This suggests the same sensitivity of these categories of passengers to changes in the basic value of the tariff (single full fare ticket). This fact allows to conclude that the distribution of the shares of tickets sold on each of the routes between these categories will remain unchanged when the full fare is changed by a certain amount while the discount rate from the full fare for these categories of passengers remains unchanged. Then we can further consider that the problem of managing revenues can be reduced to the problem of choosing the full fare, which influences both the demand of categories of passengers traveling under the full fare and the demand of passengers with federal and regional discounts.\par
It should be underlined that there is an inelastic (price insensitive) demand for such categories of tickets as children's tickets and tickets for the PPK and RR employees. The demand for these categories in the vicinity of the existing tariff values is price insensitive due to the fact that the value of the tariff for children is insignificant relatively to the cost of a full ticket (20\% to full fare). In addition, parents who take their children on local rail trips do not have the ability to change the frequency of travel for children while maintaining their frequency of travel. These two facts lead to the insensitivity of the demand for tickets for children to the price of the base fare. Employees of the PPK and RR also receive a substantial discount (up to 90\%) from the full fare. Therefore, the relatively low level of prices for this category of passengers, as well as the need for regular trips for the working duties, lead to insensitivity of demand to the value of the full fare. Due to the strict correlation of the tariffs for PPK employees with the full fare, it can also be argued that the problem of setting tariffs for all groups of passengers is reduced to the problem of setting the full fare. At the same time, it is possible to predict the change in revenue on tickets sold to the PPK employees, commensurate with the change in full due to the lack of sensitivity of the demand of this category of passengers to changes in tariff.\par
Different specifications of the demand Eq. (1) estimated by the least squares method for each of the subsamples corresponding to one of the categories of fare for tickets sold allow to obtain price sensitivity estimates that are the same for all tickets sold within the ticket category. For example, for the specification (IV) of equation (1), the estimate of the price elasticity of demand is equal to -1.981. This estimate shows the price elasticity of demand for an average trip in any of the directions and any pair of stations at the full fare for single ticket. In the meantime, price sensitivity of demand may vary depending on the magnitude of demand (“low” and “high” demand, for example, in winter and summer, in daytime and morning hours, on weekdays and weekends, etc.), degree of competition (availability of and proximity of the local bus and the highway), the typical goal of the trip (a trip to work, to the garden, or for recreation), the distance of the trip, etc. Thus, the average estimate of the sensitivity of demand to price is not an exact approximation of the real sensitivity of demand for certain categories of travel. To take into account the differences in the sensitivity of demand to price and other factors, we estimate a demand model based on the bootstrap aggregation of regression trees. The construction of a single regression tree implies an algorithmic sequential partitioning of a sample of data into subsamples by the values of explanatory variables.\par 
A model similar to Eq. (1) is calibrated for each subsample of data. The optimal partitioning into subsamples is carried out according to the criterion of the accuracy of the final prediction (the minimum mean square error of the prediction). The accuracy of the final prediction is considered on the basis of cross-validation, i.e. model parameters are calibrated on a random 75\% (training) subsample of observations, and based on the obtained parameter estimates, a prediction is made for the remaining 25\% (testing) subsample. The criterion for stopping of tree growth is the criterion of the maximum number of partitions not exceeding the cubic root of the total number of observations, as well as achieving the minimum subsample size, not less than the square of the number of parameters calibrated in the subsample (Breiman, 1996). Using partitions for various variables allows to simulate the cross effects of different variables on each other, for example, to estimate the change in price effect for different directions of travel, stations with different characteristics, etc.\par
To reduce the variability of the model we use the principle of bootstrap aggregation of predictions (bagging prediction). We train a set (2000) of trees where each tree is trained on a random 75\% subsample of observations. Thus, each observation will have expectedly 500 ((100\% - 75\%) $\times$ 2000) different predictions, each of which is obtained from a model calibrated not on the observations from which the prediction is made. This approach allows to prevent the model from overfitting to the data, and also allows to reduce the variability of the model and its sensitivity to a specific dataset and influential observations. \par
When calibrating a regression tree model, a set of individual parameter values is calculated (for each final subsample within one tree and for each tree), the estimation results are nontrivially interpretable. To study the contribution of each variable to the model we use an indicator of the “relative importance” of the variable which is calculated as the share of the sample partitions into subsamples by a given variable among all the partitions. Such an indicator reflects the contribution of the variable to the explained heterogeneity of the observations, i.e. the relative contribution of the variable in explaining the differences between the various trips. \par
The results of the importance assessment are presented in Table 7. We supress variables with the importance indicator less than 3\%.The model of regression trees in comparison with the ordinary least squares shows a significantly higher coefficient of determination (0.531) obtained on cross-validation, i.e. derived from predictions on test samples. Thus, the model explains more than a half of the total variation in demand, which is a very high number for large volume data.\par
	\begin{table}%
	\caption{Relative variable importance}
	\label{tab:07}
	\Tablefontsm
	\begin{minipage}{\columnwidth}
		\begin{center}
	\begin{tabular}{ L{7.2 cm} C{3.6 cm}}
\noalign{\smallskip}\hline\hline\noalign{\smallskip}

 Variable & Relative importance, \% \\

				\noalign{\smallskip}\hline\noalign{\smallskip}
Direction: Western	& 3.7\\
Direction: Gornozavodsk		& 3.2\\
Trip: Distance	& 	3.1\\
Departure: from Perm	& 	7.5\\
Departure: summer gardens 	& 	5.6\\
Departure: population from 100 to 1000	& 	5.9\\
Departure: geographic longitude	& 	3.4\\
Departure: distance to the closest settlement, km	& 	5.3\\
Departure: distance to the 2nd closest settlement, km	& 	3.7\\
Departure: distance to the closest highway, km	& 	3.9\\
Arrival: geographic longitude	& 	5.0\\
Arrival: distance to the closest settlement, km		& 5.9\\
Arrival: distance to the 2nd closest settlement, km		& 6.2\\
Arrival: distance to the closest highway, km	& 	3.4\\
Arrival: population up to 100	& 	4.0\\
Arrival: population		& 10.6\\
Other	& 	19.6\\

\noalign{\smallskip}\hline\noalign{\smallskip}
$CV$ $R^2$  &	0.531\\
Number of trees & 2000\\
Number of partitions & 65331\\
\noalign{\smallskip}\hline\noalign{\smallskip}

		\end{tabular}
		\end{center}
	\end{minipage}
\end{table}

Results for variables importance show that the sample is frequently divided (demand statistically differs) by the Western and Gornozavodsk directions, trip distance and characteristics of the arrival and departure stations. Among the characteristics of the stations, the most important variables are the population of the station of arrival, the size of the settlements of the stations of arrival and departure, the departure of the trip from Perm, and the location relative to the settlements of the station of arrival and departure.\par
The effects of factors explaining demand, including the sensitivity of demand to price, may have a significant variation depending on the factors described above. In order to assess the variation in the sensitivity of demand to price we numerically estimate an effect of price comparing the demand predictions with the actual price and price perturbed on 10\%. Fig.1 presents a histogram and a kernel density estimate of the empirical distribution of the price elasticity of demand over the entire sample of observations. Fig. 1 shows a significant variation in the sensitivity (elasticity) of demand to price. The average value of elasticity corresponds to the estimate of the elasticity of the average trip according to OLS. At the same time, a large proportion of observations demonstrate elastic demand (the value of elasticity in absolute value is above 1), but there is also a significant proportion of observations with weakly elastic (elasticity value in absolute value is less than 1) and inelastic (elasticity value is 0) demand. For 75\% of observations, demand is elastic, which indicates the need to reduce the tariff for this group of trips to increase total revenue while about a quarter of trips show weakly elastic or inelastic demand, which suggests that for the tariff should be increased in order to maximize the revenue.\par
\begin{figure}
    \centering
    \includegraphics[width=10cm]{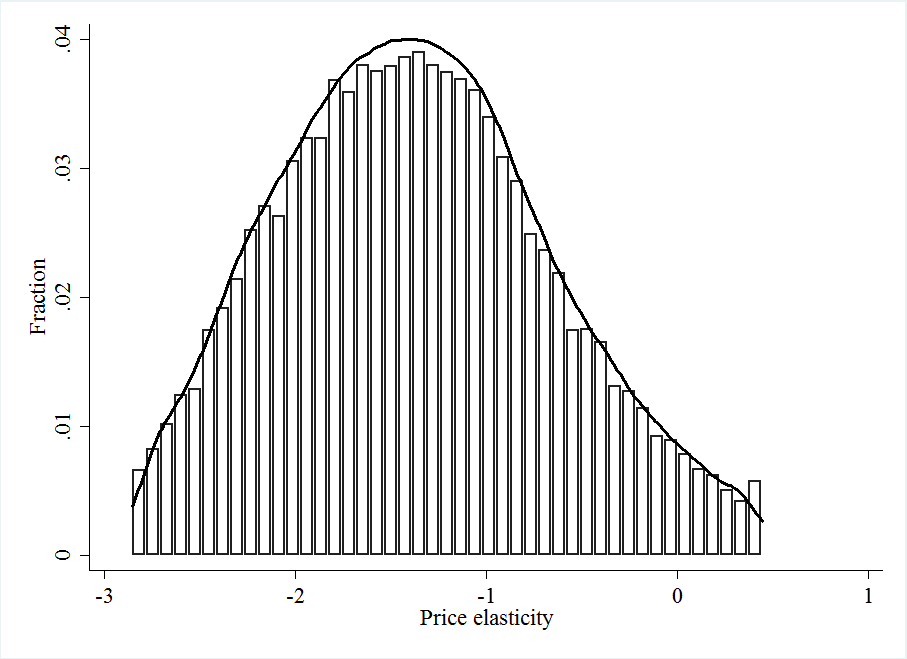}
    \caption{Distribution of price elasticity across various trips}
    \label{fig:01}
\end{figure}

Further problem is to find and describe groups of trips for which demand is elastic and weakly elastic, respectively. We group trips by directions, distances and belonging to one of the ends of a trip in Perm  since these variables are important in explaining the demand heterogeneity. The selection of cutoff points for the trip distance was made on the basis of the presence of a significant share of trips in each of the groups. Estimates of the average elasticity of trips by groups are presented in Table 8. \par
Table 8 shows that the Western direction and, in particular, its part that is not related to trips inside Perm, is the most elastic part of total demand along with trips from / to Perm in the Kungur direction. The distance of trips in the Western direction does not have a significant effect on the estimate of price elasticity, as well as the fact that trips are not to or from Perm. At the same time, in the Kungur direction, the demand for travel from Perm is more elastic the shorter the trip, which confirms the hypothesis that there is higher degree of competition in the Kungur direction with local bus transportation when passengers are traveling to summer gardens. The demand on the Kungur direction in areas outside Perm has a relatively low elasticity, close to 1 in absolute value, which may be due to the relatively poor connection between small settlements along the way and higher convenience of local bus services compared to railway transport.\par
	\begin{table}%
	\caption{Estimates of price elasticity for various trip directions and types}
	\label{tab:08}
	\Tablefontsm
	\begin{minipage}{\columnwidth}
	\begin{center}
	\begin{tabular}{L{0.1cm} L{2.6 cm} C{1.7 cm} C{1.7 cm} C{1.7 cm} C{1.9 cm} C{1.7 cm} }
\noalign{\smallskip}\hline\noalign{\smallskip}

& & Western & Kungur & \multicolumn{2}{c}{Gornozavodsk} & Agglom.\\  
\multicolumn{2}{l}{Type of trip} & & & Perm-Chus., & Chus. - Un',\\
& & & & Perm-Kizel & Chus. - Evrop.\\
\noalign{\smallskip}\hline\noalign{\smallskip}
\multicolumn{2}{l}{All types} &	-1.75 &		-1.34 &		-1.26 &		-0.89 &		-1.21\\
\noalign{\smallskip}\hline\noalign{\smallskip}
\multicolumn{2}{l}{From/to Perm} &		-1.72 &		-1.69 &		-1.41 &		-	 &	-\\
 &     Short (1-4 zones) &	 	-1.71 &		-1.92 &		-1.54 &		-	 &	-\\
 &     Middle (5-8 zones) &	 	-1.73 &		-1.77 &		-1.49 &		- &		-\\
 &     Long (9-17 zones) &	 	-1.72 &		-1.55 &		-1.26 &		-	 &	-\\
\noalign{\smallskip}\hline\noalign{\smallskip}
  \multicolumn{2}{l}{Out of Perm} &		-1.79 &		-1.18 &		-0.93 &		-0.89 &		-\\
   &	    Short (1-3 zones)  &		-1.78 &		-1.19 &		-0.96 &		-0.87 &		-\\
    &	  Middle (4-6 zones) &	 	-1.73 &		-1.36 &		-0.98 &		-0.95 &		-\\
 &	  Long (7-17 zones) &	 	-1.85 &		-1.05 &		-0.89 &		-0.69 &		- \\
\noalign{\smallskip}\hline\noalign{\smallskip}
   \multicolumn{2}{l}{Within Perm} &	 -1.20 &	 -1.72 &	 -1.19 &	 - &	 -1.21\\
 &	Short (1 zones) &	 	-1.12 &		-1.72 &		-1.17 &		- &		-1.17\\
 &	 Middle (2 zones) &	 	-1.68 &		-	 &	-1.28 &		- &		-1.34\\
\noalign{\smallskip}\hline\noalign{\smallskip}
	\end{tabular}
	\end{center}
	\end{minipage}
\end{table}

Also elastic but so far from unit elasticity demand is observed for trips within the urban agglomeration. Thus, the price of a ticket for a railway trip within the agglomeration was set higher than the cost of travel by public transport until 2016, as a result of which a reduction in the tariff should predictably lead to an increase in revenue from sold tickets. Weakly elastic demand is observed for trips of various distances along the Gornozavodsk directions along routes that do not include Perm, including the part from Chusovoy to Europeyskaya and Un'. Weakly elastic demand here is implied from using rail transport with a relatively low level of development of alternative modes of transport. Also, the presence of the typical purpose of the trip inflexible frequency  with an increase in the price, for example, to work, may also indicate a weak elasticity of demand. This trip purpose is typical as travel between a large settlement with a large number of jobs and good district infrastructure (Chusovoy, Lys'va, Ugleuralsky, Gornozavodsk) and a smaller settlement that is the place of residence. The demand for trips from / to Perm in the Gornozavodsk direction is elastic both for short-range trips (to summer gardens, for example, Palniki, Div'ya, Yarino), and for medium- and long-distance trips (in Chusovoy, Ugleuralsky, Lys'va). Thus, the analysis of the sensitivity of demand to price in terms of trip goals, trip directions and distance shows the directions of possible  change in tariffs for certain types of trips to optimize the company revenue. Thus, the demand for an average trip is elastic, which means that the average rate of tariff growth should not outpace the growth of income, which will ensure a reduction in the real tariff and revenue growth. At the same time, the tariff should be differentiated in various directions and trips of various distances. Thus, the presence of weakly elastic demand for some parts of the Gornozavodsk direction indicates the need to increase the real tariff for this category of trips. Travel types with demand elasticity in absolute value slightly exceeds 1 (trips in agglomeration, trips in Kungur direction outside Perm) require indexation of tariff from prices of 2016 by less than the value of income growth. This will reduce the real tariff and increase revenue. Highly elastic demand (all the Western direction and trips from Perm to Gornozavodsk and Kungur directions) require a substantial reduction in the real tariff to increase the revenue. Thus, these areas are characterized by high competition with other modes of transport and the flexible possibilities of passengers to change the frequency of travel by rail, which requires lower prices to increase passenger traffic and revenue.

\section{Conclusion}

In this paper we estimate the demand function for the purpose of further optimizing the tariff system for local rail company in the Perm Territory, Russia. The analysis of the current tariff policy allows to conclude about the heterogeneity of trips in various travel directions, differences in typical travel objectives for certain pairs of stations, which are not caused by differences solely in the travel distance, but rather related to the specific location of settlements along travel directions and characteristics of populated areas between which the trip is carried out. These differences in travel demand structure are caused by differences in the market size of each individual trip and differences in the sensitivity of demand to a potential tariff change. According to the current pricing system of the PPK, tariffs are set only on the basis of travel distance. At the same time trips of the same distance range may occur to be trips with completely different demand patterns mixed into one tariff zone. It may mix up both trips between small settlements, trips to work to another settlement, trips to the summer garden, and trips for the weekend for the purpose of having a rest, etc. Consequently, the demand should be analyzed for each individual pair of stations taking into account the potential unobservable demand heterogeneity at the level of each individual trip route, characterized by the trip distance, purpose and type of trip, trip direction.\par
In this paper we employ actual data on tickets sales by the PPK for 2012-2016, obtained from the internal ticket sales system. Sales data was combined with PPK data on actual train schedule for the same period. We also collect a set of data, containing the characteristics of each station in the PPK area of operation, that include information about the settlements closest to the station, stops of local buses and highways. \par
We aggregate sales to the monthly level for each pair of stations and construct a demand prediction model where sales are determined by the value of the tariff, the characteristics of the trip and the stations. We assume characteristics of a trip and stations as potential sources of demand heterogeneity. The the model of the bootstrap aggregation of regression trees allow to estimate the market demand capacity and the sensitivity of demand to tariff change separately for each route (trips between a particular pair of stations). \par
We reveal that demand is elastic by price in average, while the estimate of demand elasticity significantly varies over trips. A quarter of trips has a weakly elastic demand, which is mainly due to the lack of close substitutes for railway transport in certain areas and the inability of passengers to change the frequency of trips due to specific typical travel goals. As a result, one may conclude that for a number of trips that have elastic demand, the optimal pricing strategy will be to reduce the real value of the tariff, while increasing the revenue from trips with weakly elastic demand requires an increase in the real tariff.\par
A number of technological and institutional constraints limits the performed analysis. Limitations associated with the lack of variation in some variables in a dataset in the period of observation do not allow to identify differences in demand in some dimensions, and also subsequently differentiate the tariff in some dimensions. Thus, a key technological and data limitation is the lack of a ticket validation system. This leads to the fact that the actual train on which the trip was made and the actual train fullness is unknown. Coupled with the institutional restriction on the possibility of travel at any time of the day resulted in possibility of ticket purchase for a particular day, it is impossible to differentiate the fare by date and time of a travel (set a different fare for different train arrival time, for trains on different days within a week and months within a year). However, it is known from the literature on the pricing management in railway industry that this kind of price discrimination over a demand periods also allows to increase the total revenue from ticket sales. Another important limitation is the lack of variation in the amount of discounts on quantity. Thus, the cost of a round trip ticket, as well as the discount for season tickets, remained constant during the period in observation, which does not allow to simulate the passengers' choice of a ticket type and analyze the optimal value for such a discount. However, the discount rate for various ticket types is also an important problem for differentiating the fare. Being able to model switching behavior between different ticket types, would also lead to an additional source of price discrimination and revenue increase. In addition, the lack of validation of tickets does not allow to sell a ticket with reference to a specific place. In general, managing the assortment of tickets, such as the possibility of buying a ticket with and without reference to a place, selling tickets to coaches with different degree of comfort and selling other additional comfort-enhancing services, could also help increase the company's total revenue from ticket sales with wider product range.
\section*{References}

\begin{enumerate}

\item Bajari, P., Nekipelov, D., Ryan, S. P.,  Yang, M. (2015). Machine learning methods for demand estimation. The American Economic Review, 105(5), 481-485.
\item 	Ben-Akiva, M. and Lerman, S. (1985) Discrete Choice Analysis: Theory and Applications to Travel Demand. Cambridge, MA: MIT Press.
\item 	Bhat, C. R., and Castelar, S. (2002). A unified mixed logit framework for modeling revealed and stated preferences: formulation and application to congestion pricing analysis in the San Francisco Bay area. Transportation Research Part B: Methodological, 36(7), 593-616.
\item 	Breiman, L. (1996). Bagging predictors. Machine learning, 23(2), 124-140.
\item 	Brownstone, D., Bunch, D. and Train, K. (2000). Joint mixed logit models of stated and revealed preferences for alternative-fuel vehicles. Transportation Research Part B 34(5): 315–338.
\item 	Carrier, E. (2003) Modeling airline passenger choice: Passenger preference for schedule in the passenger origin – Destination simulator (PODS). MS thesis, Massachusetts Institute of Technology.
\item 	Chaneton, J. and Vulcano, G. (2011) Computing bid-prices for revenue management under customer choice behavior. Manufacturing Service Operations Management 13(4): 452–470.
\item 	Cherchi, E. and Ortúzar, J.de.D. (2003) Alternative specific variables in non-linear utility functions: Influence of correlation, homoscedasticity and taste variations. 10th International Conference on Travel Behaviour Research, Lucerne, Switzerland.
\item 	Greene, W.H., Hensher, D.A. and Rose, J. (2006) Accounting for heterogeneity in the variance of unobserved effects in mixed logit models. Transportation Research Part B 40(1): 75–92.
\item 	Hess, S. and Polak, J. (2005) Mixed logit modelling of airport choice in multi-airport regions. Journal of Air Transport Management 11(2): 59–68.
\item 	Hess, S., Train, K. and Polak, J. (2006) On the use of modified latin hypercube sampling (MLHS) method in the estimation of mixed logit model for vehicle choice. Transportation Research Part B 40(2): 147–163.
\item 	Hetrakul, P. and Cirillo, C. (2013) Accommodating taste heterogeneity in railway passenger choice models based on internet booking data. Journal of Choice Modeling 6(C): 1–16.
\item 	Ly, T. (2012) Railway industry in France. Revenue management in action.\\ http://blogs.cornell.edu/advancedrevenuemanagement12/2012/03/26/\\railway-industry-in-france/.
\item 	Morikawa, T., Ben-Akiva, M. and McFadden, D. (2002). Discrete choice models incorporating revealed preferences and psychometric data. In: Advances in Econometrics (pp. 29-55). Emerald Group Publishing Limited.
\item 	Newman, J., Ferguson, M. and Garrow, L. (2012) Estimating discrete choice models with incomplete data. Transportation Research Record 2302: 130–137.
\item 	Rietveld, P. (2000). The accessibility of railway stations: the role of the bicycle in The Netherlands. Transportation Research Part D: Transport and Environment, 5(1), 71-75.
\item 	Talluri, K. and van Ryzin, G.J. (2004a) Revenue management under a general discrete choice model of consumer behavior. Management Science 50(1): 15–33.
\item 	Talluri, K. and van Ryzin, G.J. (2004b). The Theory and Practice of Revenue Management. New York: Kluwer Academic.
\item 	Vulcano, G., van Ryzin, G.J. and Chaar, W. (2010). Choice-based revenue management: A study of estimation and optimization. Manufacturing  Service Operations Management  12(3): 371–392.
\item 	Van Vuuren, D. (2002). Optimal pricing in railway passenger transport: theory and practice in The Netherlands. Transport policy, 9(2), 95-106.
\item 	Van Vuuren, D.,  Rietveld, P. (2002). The off-peak demand for train kilometres and train tickets: A microeconometric analysis. Journal of Transport Economics and Policy (JTEP), 36(1), 49-72.
\item 	Wardman, M., Lythgoe, W.,  Whelan, G. (2007). Rail passenger demand forecasting: cross-sectional models revisited. Research in transportation economics, 20, 119-152.
\item 	Whelan, G., Batley, R., Shires, J.,  Wardman, M. (2008). Optimal fares regulation for Britain’s railways. Transportation Research Part E: Logistics and Transportation Review, 44(5), 807-819.
\item	Whelan, G.,  Johnson, D. (2004). Modelling the impact of alternative fare structures on train overcrowding. International journal of transport management, 2(1), 51-58.
\item 	Zemp, S., Stauffacher, M., Lang, D. J.,  Scholz, R. W. (2011). Classifying railway stations for strategic transport and land use planning: Context matters!. Journal of transport geography, 19(4), 670-679.
	\end{enumerate}


\end{document}